\documentclass[twocolumn,showpacs,preprintnumbers,amsmath,amssymb,prb]{revtex4}
\usepackage{graphicx}% Include figure files
\usepackage{dcolumn}% Align table columns on decimal point
\usepackage{bm}% bold math

\begin{document}

\title{Resistance fluctuations and Aharonov-Bohm-type oscillations in antidot arrays \\in the quantum Hall regime}%:\\with Forced Linebreak}% Force line breaks with \\

\author{Masanori Kato}
\author{Akira Endo}%
\author{Shingo Katsumoto}%
\author{Yasuhiro Iye}%
 %\email{masanori@issp.u-tokyo.ac.jp}
\affiliation{Institute for Solid State Physics, University of Tokyo, 5-1-5 Kashiwanoha, Kashiwa, Chiba 277-8581, Japan}

%\affiliation{

\date{\today}% It is always \today, today,

\begin{abstract}
Resistance fluctuation phenomenon in antidot lattices in the quantum Hall regime are studied. 
Magnetoresistance of finite antidot array systems in the quantum Hall plateau transition regime exhibits two types of oscillatory effect.
One is the aperiodic resistance fluctuations (RFs) and the other is the Aharonov-Bohm (AB)-type oscillations. 
Their dependences on the magnetic field and the gate voltage are quite distinct. 
While the aperiodic RFs are attributed to the complex evolution of the conducting network of compressible channels, the AB-type oscillations are interpreted in terms of edge states formed around individual antidots. 
The self-consistent screening effect is important for the both phenomenon, whereas, the single electron charging effect plays a minor role in the present case. 

\end{abstract}

\maketitle

The integer quantum Hall (QH) effect in bulk two-dimensional electron systems (2DESs) under a strong perpendicular magnetic field manifests itself as quantization of Hall resistivity $\rho_{xy}$ to values $h/ie^2$ 
($h$ is Planck's constant, $e$ is the charge of the electron, and $i$ is an integer) and concomitant vanishing of the longitudinal resistivity $\rho_{xx}$.\cite{Prange}
In the magnetic field range of transition from one integer to the next, $\rho_{xy}$ undergoes a stepwise change and $\rho_{xx}$ shows a peak.
As the temperature is lowered, the width of this so-called QH transition region decreases according to a power law $\Delta B \propto T^\kappa$ with $\kappa \sim 0.42$.\cite{Huckestein}
In contrast to the featureless steps and peaks for the case of bulk 2DES, mesoscopic 2DES samples such as quantum wires exhibit broader QH transitions accompanied with reproducible resistance fluctuations. 
Although such resistance fluctuations (RFs) are rather common in mesoscopic structures, the detailed origin of the RF phenomena in the QH regime is not fully understood.
Various models invoking resonant tunneling through a small number of localized states,\cite{Simmons} charging effect,\cite{Cobden} and the network of compressible stripe\cite{Machida} have been proposed. 
These experiments were carried out by use of quantum wire or similar mesoscopic scale samples.
The localized states, which mediate the scattering from the edge channels on one side to the opposite, are thought to be formed in the random potential landscape.
The existence (or otherwise) of the relevant localized states and their configuration are unpredictable, which obscures the interpretation of the RFs.
In this work, we use a small array of antidots, in which at least one kind of the localized states, {\it i.e.}, the edge states formed around the antidots, are well-defined.

In the antidot system, which is a regular array of artificial potentials, we have observed so-called Aharonov-Bohm (AB)-type oscillations in the quantum Hall regime. 
The high field AB-type oscillation (HFABO) is $B$-periodic with the period corresponding to one flux quantum per the antidot area. Previous experiments 
\cite {Iyeantidot, Katophysica, nu2, KatoICPS} suggest that the origin of the HFABO is explained by the single particle picture around an antidot. 

In this paper we studied the resistance oscillations in the QH regime in a small array of antidots. 
We observed the HFABOs superimposed on the RFs. To elucidate the difference of these two kinds of the oscillations we studied the evolution of the oscillations in 
the magnetic field-gate voltage plane. Our result is that the features of the RFs and HFABOs with the gate voltage show the different magnetic field dependence. 
We conclude that the HFABOs can be explained by the single-particle states around an antidot potential but the RFs cannot as proposed in the previous experiments.\cite{Cobden,Machida}

\begin{figure}[htb]
\begin{center}
\includegraphics[width=0.95\linewidth]{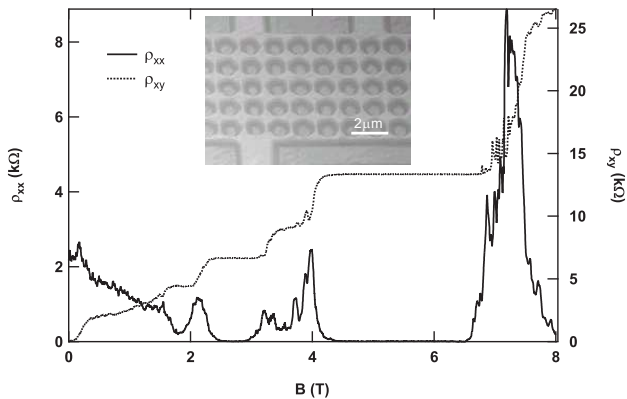}
\caption{Magnetoresistivity and Hall resistivity of a $5 \times 5$ antidot array at $V_g = 0~\mathrm{mV}$. The inset is a scanning electron micrograph of the sample.}
\label{FIG1}
\end{center}
\end{figure}

Antidot array samples were fabricated from a GaAs/AlGaAs single heterojunction wafer (density $n=4.0\times 10^{11}~\mathrm{cm^{-2}}$ and mobility $\mu =98~\mathrm{m^2/Vs}$). 
The 2DES plane was located 60~nm below the surface. 
Fabrication of the antidot array and the Hall bar shape was done by electron beam lithography and shallow (30 nm) wet chemical etching. 
The sample used in this study was a $5.3~\mathrm{\mu m}$ wide Hall bar containing $5 \times 5$ antidots in a square lattice pattern as shown in the inset of Fig.~\ref{FIG1}.
The lattice period was $a = 1~\mathrm{\mu m}$ and the antidot radius was $r = 350~\mathrm{nm}$. 
The effective radius $r^*$ of the antidots was larger than the lithographical radius $r$ typically by $\sim 100~\mathrm{nm}$, so that the channels between two neighboring antidots were $\sim 100~\mathrm{nm}$ wide at their narrowest point. 
This sort of antidot array system can be viewed as a network of a narrow conducting channels. 
The ohmic contacts to the Hall bar were made with AuGe/Ni electrode pads. 
A Au-Ti Schottky front gate enabled us to tune the Fermi energy of the system. 
The sample was cooled down to $30~\mathrm{mK}$ in a mixing chamber of a top-loading dilution refrigerator. 
The longitudinal and Hall resistances were measured as a function of the perpendicular magnetic field $B$ and the gate bias $V_g$ by a standard low frequency (13 Hz) ac lock-in technique with an excitation current of 1 nA. 
The experimental results presented here were obtained in two cool-down cycles.

Figure~\ref{FIG1} shows the traces of $\rho_{xx}$ and $\rho_{xy}$ for zero gate bias ($V_g = 0~\mathrm{mV}$) as a function of perpendicular magnetic field $B$.
In the vicinity of zero magnetic field, characteristic features of the antidot lattice, {\it e.g.}, a commensurability peak of $\rho_{xx}$ and quenching of the $\rho_{xy}$, are seen.
Unlike bulk 2DES samples, the QH plateau transitions exhibit rich structures, as is characteristic to mesoscopic samples.

\begin{figure}[htb]
\begin{center}
\includegraphics[width=0.95\linewidth]{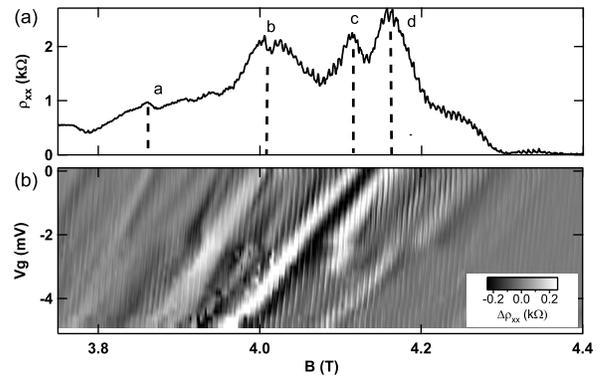}
\caption{(a) $\rho _{xx}$ and $\rho _{xy}$ at $V_g=0~\mathrm{mV}$ in the filling factor range between $\nu $=3 and 2. 
(b) Gray-scale plot of $\Delta \rho _{xx}$ in sweep of $B$ and $V_g$. Peak {\bf a},{\bf b},{\bf c},{\bf d} in (a) correspond to white lines in (b).}
\label{FIG2}
\end{center}
\end{figure}

Figure \ref{FIG2}(a) shows the details of $\rho_{xx}$ in the QH transition region between $\nu = 2$ and 3.
The following two features are conspicuous.
One is the rapid periodic oscillations clearly visible on the higher $B$ side of the transition ($3.9<B<4.4$ T).
The period 6.5 mT corresponds to the AB period $\Delta B=h/eS$ expected for the effective antidot area $S=\pi {r^*}^2$ with $r^*=450$ nm.
The other feature is the aperiodic RFs which are responsible for the rich structures of the QH transition peaks seen in Fig.~\ref{FIG1}. 

These two features, namely HFABO and RF, evolve differently with the gate bias.
This is seen in Fig.~\ref{FIG2}(b) which shows the resistance variation as a function of $B$ and $V_g$.
Here, $\Delta \rho_{xx}$ plotted in gray scale represents the local variation of the longitudinal resistance after subtracting a slowly varying background.
The light and dark regions represent local peaks and dips of $\rho_{xx}$, respectively. 
The data shown in Fig.~\ref{FIG2}(a) correspond to the top edge of this figure.
It is evident from this figure that the trajectories of the HFABO and RF in the $B$-$V_g$ plane have quite distinct slopes.
The fine stripes of the HFABOs are only slightly slanted from the vertical.
The slope expressed as $\Delta B/\Delta V_g$ is about $4~\mathrm{T/V}$ in this region.
By contrast, the stripes of the RFs that correspond to the peaks labelled {\bf a} to {\bf d} in Fig.~\ref{FIG2}(a) are much less steep. 
Moreover, the slope of the stripes {\bf a} and {\bf b} on the higher $B$ side of the QH transition is clearly different from that of the stripes {\bf c} and {\bf d} on the lower $B$ side. 

\begin{figure}[htb]
\begin{center}
\includegraphics[width=0.95\linewidth]{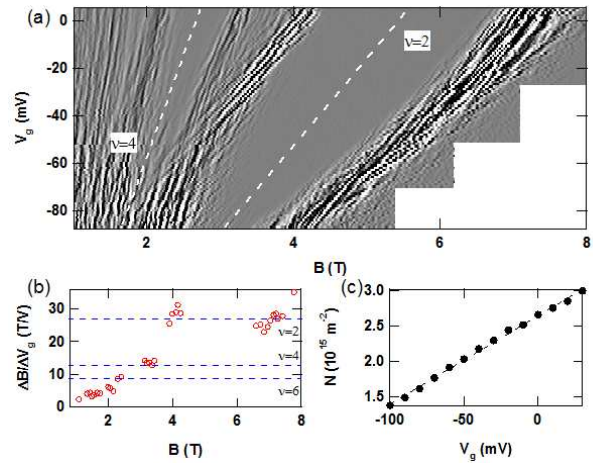}
\caption{(Color online) (a) Gray-scale plot of second-order differential resistivity in sweep of $B$ and $V_g$. The white dashed lines 
represent integer filling of Landau level. 
(b) The slopes of resistance fluctuations at $V_g=0~\mathrm{mV}$. 
The (blue) dashed lines indicate the value of the slope calculated by Eq.(\ref{eq:BVL}).
(c) The dependence of density on the gate voltage $V_g$ in the antidot sample.}
\label{FIG3}
\end{center}
\end{figure}

The gray scale plot in Fig.~\ref {FIG3}(a) shows the global evolution of the RFs over a wider region of the $B$-$V_g$ plane.
Here, the second derivative of the longitudinal resistance $\partial^2 \rho _{xx}/\partial B^2$ is plotted in gray scale so as to highlight the RFs.
(The HFABOs are too fine to be resolved on this scale.)
The trajectories of the RF extrema in this plot are seen to be approximately straight lines whose slopes depend on the filling range.
The two dashed lines drawn in Fig.~\ref{FIG3}(a) represent the positions of $\nu = 4$ and 2.
It is seen that the RF extrema on either side of the QH state move parallel to the dashed lines.
Similar behavior is also observed at higher values of $\nu$.
This implies that the trajectories of the RF extrema in the vicinity of the integer LL filling $\nu$ can be represented as
\begin{equation}
\frac {\Delta B}{\Delta V_g}= \frac{1}{\nu } \frac {hC}{e^2},
\label{eq:BVL}
\end{equation}
\noindent
where $C$ is the capacitance between the gate and the 2DES. 
The above relation is derived from $\Delta n/\Delta V_g=C/e$, and $\Delta n/\Delta B=\nu e/h$. 
The value of $C/e$ can be obtained from the $n$ versus $V_g$ relation shown in Fig.~\ref {FIG3}(c), which was obtained from the analysis of the magnetic field positions 
for the Hall plateaus as a function of the gate voltage. 

Figure~\ref {FIG3}(b) shows the values $\Delta B/\Delta V_g$ at $V_g = 0~\mathrm{mV}$ in the magnetic field intervals where the RFs are visible. 
The horizontal dashed lines represent the values of $\Delta B/\Delta V_g$ obtained from Eq.~(\ref{eq:BVL}) with $\nu = 2$, 4 and 6.
It is seen that $\Delta B/\Delta V_g$ takes the same value on either side of the $\nu = 2$ QH state, as indicated by the bands of stripes parallel to the dashed line of constant filling $\nu = 2$ in Fig.~\ref {FIG3}(a). 
Similar bahavior is also observed for the RFs on the higher field flank of the $\nu = 4$ minimum.
The evolution of the RFs with $V_g$ becomes more difficult to track down at lower fields ({\it i.e.}, for higher values of $\nu$), but the qualitative trend is the same.

Similar observations have been reported by Cobden {\it et al.}\cite{Cobden} for mesoscopic Si MOSFET samples and by Machida {\it et al.}\cite{Machida} for GaAs/AlGaAs heterostructures. 
Cobden {\it et al.}\cite{Cobden} pointed out that this behavior cannot be explained in terms of the non-interacting model\cite{Jain} which associates 
the RFs with successive alignment of the Fermi level with scattering resonances in particular Landau bands, and suggested the importance of the Coulomb charging effect.
They argued that the pattern of the incompressible strips surrounding the compressible (metallic) regions should be preserved along the lines parallel to the constant integer fillings 
so that the charging condition remains unchanged.

Machida {\it et al.}\cite{Machida} studied a system consisting of two QH transition regions coherently connected in series by the QH edge channels and found the RF pattern to obey simple addition of the RFs of the two individual regions.
They also noted that the typical fluctuation period in $V_g$ was about a hundred times as large as the average level spacing and argued that the single electron charging effect should be unimportant. 
Based on these observations, they have concluded that the RF phenomenon is classical in origin, and have proposed a model based on a network of compressible (conducting) regions whose pattern is supposed to be sensitive to the change in the electron density.

It should be noted, however, that the smallness of the average level spacing compared with the energy scale of the fluctuation does not by itself preclude the charging effect.
This is readily seen if one considers the case of metallic quantum dots where average level spacing is very small. 
The energy scale of the single electron charging effect is determined by the capacitance of the conductance-limiting coulomb island.
The conditions for the single electron charging effect to be dominant are that such coulomb islands with sufficiently large charging energy exist in the system, and that one (or very few) of them acts as conduction bottleneck ({\it i.e.}, not by-passed by other conducting paths).
Whether these conditions are met or not depends on the particular system.
They may be met in some of the small Si MOSFET samples of Cobden {\it et al.}'s work,\cite{Cobden} and probably not in the GaAs samples with a larger overall length studied by Machida {\it et al.}\cite{Machida}
Therefore it is possible that their models can be both right for their respective situations.

As for the present system, we conclude that the single electron charging is unlikely to be the principal cause of the RFs for the following reasons.
First, the overall size of our system is not too small.
Secondly, because the RFs are observed in the QH transition regime where conductance is relatively high, it is unlikely that the conduction is limited just by one or very few bottlenecks.
Thirdly, no (quasi-)periodic oscillation pattern as a function of $V_g$ is observed for the RFs.
(Regular periodicity would be a signature of the single electron charging effect in a system with small average level spacing.) 

Let us now turn to the features of the HFABOs.
The trajectories of the HFABO extrema in Fig.~\ref{FIG2}(b) are much steeper than those of the RFs. 
The periodicity in $B$ is determined by the effective antidot area as stated earlier.
The periodicity in $V_g$ is about $\Delta V_g \approx 2~\mathrm{mV}$ in Fig.~\ref{FIG2}(b). 
As discussed in earlier publications,\cite{Iyeantidot, Katophysica, nu2} the HFABO is attributed to successive crossing through the Fermi level, of the single particle (SP) states formed around the antidot.
Each SP state encloses an integer number of flux quanta. 

\begin{equation}
\pi r_m^2 B = m \frac {h}{e} ~~~~~~~\mathrm {(\mathit {m}: integer)}.   
\label{eq:BS}
\end{equation}

When the external magnetic field is increased, the radius $r_m$ decreases so as to keep the quantization condition, Eq.~(\ref{eq:BS}), until the energy of the SP state rises above the Fermi level.
The difference in radius of the successive SP states is obtained as

\begin{eqnarray}
\Delta r_m &=& r_{m+1} - r_m  \cr
&\simeq & \frac {1}{2 \pi r_m B} \frac{h}{e} \cr
&=& \frac{\ell_B^2}{r_m},  
\label{eq:dr}
\end{eqnarray}

\noindent
where $\ell_B=(\hbar/eB)^{1/2}$ is the magnetic length.

The HFABOs also occur when the gate bias is swept so as to change the chemical potential (Fermi energy) of the 2DES. 
Each stripe of the HFABO in Fig.~\ref{FIG2}(b) corresponds to the trajectory of constant $m$ in Eq.~(\ref{eq:BS}), which is given by the following relation obtained by differentiating Eq.~(\ref{eq:BS}).

\begin{eqnarray}
\frac {\Delta  B}{\Delta V_g} &=& - \frac {2B}{r_m} \left(\frac {\Delta r_m}{\Delta V_g}\right) \cr
&=& - \frac{2B}{r_m} \left( \frac{d r_m}{d E} \right)_{E_\mathrm{F}} \frac{\Delta E_\mathrm{F}}{\Delta V_g}
\label{eq:dB}
\end{eqnarray}

\noindent
Here, $\Delta r_m/\Delta V_g$ denotes the change by the gate bias $V_g$, of the radius $r_m$ of the SP state at the Fermi level enclosing $m$ magnetic flux quanta. 
In the field range of interest ($2< B < 8~\mathrm{T}$), $m$ is quite large (typically $m \approx 1000$), and the radii of successive SP states differ only by $\sim 0.3~\mathrm{nm}$. 
Thus we can safly neglect the magnetic field dependences of  $r_m$ and $\Delta r_m/\Delta V_g$ at the constant gate voltages and replace
$r_m$ in Eq.~(\ref{eq:dB}) with the effective antidot radius $r^*$ which is $\sim 450~\mathrm{nm}$ for $V_g=0~\mathrm{mV}$ as shown in the previous experiments.\cite {KatoICPS} 

\begin{figure}[htb]
\begin{center}
\includegraphics[width=0.95\linewidth]{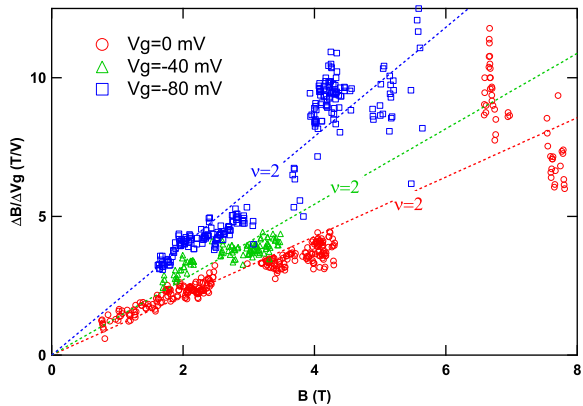}
\caption{(Color online) Red circle, green triangle and blue square show the slopes of AB oscillations $\Delta B/\Delta V_g$ obtained at 0, $-40$ and $-80$ mV. 
The dashed straight lines are linear fit as depicted in the text. }
\label{FIG4}
\end{center}
\end{figure}

Let us take the filling range between $\nu = 3$ and 2 as a representative QH transition regime.
The effective antidot radius obtained from the AB periodicity is $r^*=450~\mathrm{nm}$ for $V_g = 0~\mathrm{mV}$ and $B = 4.0~\mathrm{T}$ ($\nu = 2.7$).
From the slope of the AB stripes $\Delta B/\Delta V_g = 4.1~\mathrm{T/V}$ in Fig.~\ref{FIG2}, the rate of change in the effective radius is calculated as $\Delta r^*/\Delta V_g = -0.23~\mathrm{nm/mV}$.
If we use the value $\Delta E_\mathrm{F}/\Delta V_g = 0.046$ eV/V estimated from Fig.~\ref{FIG3}(c), this is translated to $(dE/dr)_{r^*} = -2.0 \times 10^5~\mathrm{eV/m}$.
The values of $\Delta B/\Delta V_g$ in different magnetic field ranges are shown in Fig.~\ref{FIG4} for three values of $V_g$.
The approximately $B$-linear behavior of $\Delta B/\Delta V_g$ reflects the validity of our assumption that  $r_m$ and $\Delta r_m/\Delta V_g$ in Eq.~(\ref{eq:dB}) do not change with the magnetic field $B$ largely.

As the front gate bias is made more negative, $\Delta B/\Delta V_g$ for the same magnetic field becomes larger.
With decreasing (more negative) bias voltage ({\it i.e.}, decreasing Fermi energy), the position of the edge state moves further away from the geometrical edge of the antidot because of the wider depletion from an antidot, so that the local potential gradient becomes less steep.
The corresponding values at $\nu =2.7$ for $V_g = -40~(-80)~\mathrm{mV}$ and $B = 3.2~(2.4)~\mathrm{T}$ are $r^* = 460~(470)~\mathrm{nm}$ and $(dE/dr)_{r^*} = -1.5~(-1.0)\times 10^5~\mathrm{eV/m}$, respectively.
This trend is qualitatively understood by considering the profile of the antidot potential.

The values of the potential slopes are significantly smaller than that of the {\it bare} potential at the edge of the antidots, which typically takes a value of order $10^6~\mathrm{eV/m}$ determined by the potential profile of the depletion region.
Thus the experimentally obtained values of $(dE/dr)_{r^*}$ are greatly modified (reduced) by the screening effect.
The self-consistent screening of the edge state is theoretically treated by Chklovskii {\it et al.}.\cite{Chklovskii}

In the limiting case of perfect screening, the potential in the compressible region becomes completely flat, so that the SP states at the Fermi level are highly degenerate.
In this case, the period of the conductance oscillation as a function of $V_g$ solely arise from the charging energy, as envisaged by Ford {\it et al.}\cite{Fordsingle} for the single antidot system.
If, on the other hand, the screening is partial so that the potential retains a finite slope, the period in $V_g$ reflects the SP level spacing (provided the charging energy is much smaller than it), as put forward by Karakurt {\it et al.}.\cite{Karakurt}
For the present system, the latter picture seems more appropriate.
This is inferred from the fact that the periodicity, $\Delta V_g \approx 1~\mathrm{mV}$, of the HFABO is on the same order of magnitude as that of the RFs discussed earlier, and that the single electron charging effect appears not to play a role in the latter. 
Another evidence in support of the above picture is obtained from the study of the temperature dependence (to be reported elsewhere) which has revealed the change in the $V_g$-periodicity 
qualitatively consistent with the temperature dependence of the screened edge potential as calculated by Lier and Gerhardts.\cite{Lier}

We comment on the behavior on the higher field side of $\nu =2$.
The values of $\Delta B/\Delta V_g$ exhibit significant deviation from the approximately $B$-linear behavior at the lower fillings, as seen in Fig.~\ref {FIG4}. 
The upper deviation around $B=6.7~\mathrm{T}$ for $V_g=0~\mathrm{mV}$ (red circles) implies that the slope of the antidot potential in this field range is flatter than in other field ranges. 
We note that this is the field range in which the HFABO with a \lq\lq exotic" periodicity between $h/2e$ and $h/3e$ has been observed.\cite{nu2}
The \lq\lq exotic" AB period is attributed to an orbit formed away from the edge of the antidot. 
Formation of such an orbit is more likely to occur when the edge potential becomes flatter.

In conclusion we have studied the periodic Aharonov-Bohm-type oscillations superimposed on the aperiodic resistance fluctuations in a finite antidot array in the QH plateau transition regime. 
The fluctuating resistance plotted on the $B$-$V_g$ plane exhibits stripe patterns whose slope furnishes the information on their origins. 
The aperiodic RF is attributed to the complex (and apparently irregular) evolution of the self-consistently screened potential landscape and hence the corresponding 
evolution of the network of compressible stripes in the QH transition regime. 
On the other hand, the salient features of the HFABO can be explained by the SP states associated with the edge channels formed around the antidot. 
The potential profile around the antidot at the Fermi energy can be obtained from the slope of the HFABO in the $B$-$V_g$ plane. 
The energy interval of the SP states indicates that the electrostatic potential in the compressive region is strongly modified by screening. 

\begin{acknowledgments}
This work is supported by the Grant-in-Aid for Scientific Research from the Ministry of Education, Culture, Sports, Science and Technology (MEXT), Japan, and by Special Coordination Funds for Promoting Science and Technology.
One of the authors (M.K.) gratefully acknowledges the support by Research Fellowship for Young Scientists from the Japan Society for the Promotion of Science (JSPS).
\end{acknowledgments}

\end{document}